\documentclass[runningheads]{llncs}
\usepackage[T1]{fontenc}
\usepackage{graphicx,verbatim}
\usepackage{amsfonts}
\usepackage{amsmath} 
\usepackage{subcaption}
\usepackage{cite}
\usepackage[subtle]{savetrees}

\usepackage[hidelinks]{hyperref}
\usepackage{color}

\begin{document}

\title{Physics-Based iOCT Sonification for Real-Time Interaction Awareness in Subretinal Injection}




%

\author{Luis D. Reyes Vargas\thanks{These authors contributed equally to this work.} \inst{1,5} \and Veronica Ruozzi* \inst{1,5} \and  Andrea K. M. Ross  \inst{2} \and \\ Shervin Dehghani \inst{1} \and Michael Sommersperger \inst{1} \and Koorosh Faridpooya \inst{3} \and Mohammad Ali Nasseri \inst{2} \and Merle Fairhurst \inst{4,6} \and Nassir Navab \inst{1,5} \and Sasan Matinfar \inst{1,5,6}}  

\authorrunning{L.D. Reyes Vargas et al.}
\institute{
Computer Aided Medical Procedures, Technische Universit{\"a}t M{\"u}nchen, Munich,
Germany \\
    \and TUM Klinikum Rechts der Isar, Technische Universit{\"a}t M{\"u}nchen, M{\"u}nchen, Germany\\
    \and Rotterdam Eye Hospital, Rotterdam, The Netherlands\\ 
    \and Centre for Tactile Internet with Human-in-the-Loop, Technische Universit{\"a}t Dresden, Dresden, Germany\\ 
    \and Munich Center for Machine Learning, Munich, Germany\\ 
    \and Chair for Social Affective Touch, Technische Universit{\"a}t Dresden, Dresden, Germany\\
\email{luis.reyes@tum.de}
} 
  
\maketitle              

\begin{abstract}

Subretinal injection is a delicate vitreoretinal procedure requiring precise needle placement within the subretinal space while avoiding perforation of the retinal pigment epithelium (RPE), a layer directly beneath the target with extremely limited regenerative capacity. To enhance depth perception during cannula advancement, intraoperative optical coherence tomography (iOCT) offers high-resolution cross-sectional visualization of needle–tissue interaction; however, interpreting these images requires sustained visual attention alongside the en face microscope view, thereby increasing cognitive load during critical phases and placing additional demands on the surgeon’s proprioceptive control. In this paper, we propose a structured, real-time sonification framework designed for extensible mapping of iOCT-derived anatomical features into perceptual auditory feedback. The method employs a physics-inspired acoustic model driven by segmented retinal layers from a stream of iOCT B-scans, with needle motion and injection-induced retinal layer displacements serving as excitation inputs to the sound model, enabling perception of tool position and retinal deformation. In a controlled user study (n=34), the proposed sonification achieved high retinal layer identification accuracy and robust detection of retinal deformation–related events, significantly outperforming a state-of-the-art baseline in overall event identification (83.4\% vs. 60.6\%, p < 0.001), with gains driven primarily by enhanced detection of injection-induced retinal deformation. Participants also reported higher confidence with the proposed method correlating with correctness, indicating the auditory mapping was perceptually interpretable. Evaluation by experts (n=4) confirmed the clinical relevance and potential intraoperative applicability of the method. These results establish structured iOCT sonification as a viable complementary modality for real-time surgical guidance in subretinal injection.

\end{abstract}

\section{Introduction}


Subretinal injection is a delicate vitreoretinal procedure in which a microsurgical cannula is advanced through the neurosensory retina to deliver therapeutics into the subretinal space. This approach demonstrates significant potential for application in cell- and gene-based therapies~\cite{Irigoyen2022SubretinalInjection,ling_lentiviral}. The cannula must first penetrate the Inner Limiting Membrane (ILM) before advancing within the subsequent layers for injection, which in the macular region measures only approximately $250\ \mu\mathrm{m}$ in thickness on average \cite{Jo2011Diurnal}. Immediately beneath the target lies the retinal pigment epithelium (RPE), a layer with very limited regenerative capacity where even minimal overshoot may cause irreversible damage~\cite{George2021TheRP,Zhao2017DevelopmentOA}. Because penetration forces are extremely small and masked by friction along the instrument shaft, tactile cues are largely imperceptible, forcing surgeons to rely almost exclusively on visual feedback during insertion~\cite{tactile_feedback_Gupta1999SurgicalFA}.

In clinical practice, the en face surgical microscope view serves as the primary guidance modality; however, it provides mainly surface-level information and lacks direct visualization of needle penetration depth relative to the RPE. To address this limitation, intraoperative OCT (iOCT) provides high-resolution cross-sectional views of retinal layers and tool position that are directly relevant for depth control in subretinal injection. In current workflows, iOCT is typically displayed as an overlay on the primary microscope image, requiring simultaneous interpretation of both visual streams. Large prospective studies report that iOCT influences intraoperative decision-making in only a minority of posterior-segment cases ($\approx 30\%$)~\cite{Ehlers2018DISCOVER}, suggesting that cross-sectional information may not be optimally integrated into routine surgical practice despite its clinical relevance. The addition of this secondary high-bandwidth visual stream may increase cognitive load and promote attentional tunneling during critical phases such as needle insertion~\cite{SchuetziOCTCognitive,HughesHallett2015InattentionBI}, when precise depth estimation is essential.

Prior work on OCT-based sonification has focused on epiretinal membrane (ERM) peeling~\cite{Matinfar2024OcularSA}. However, its auditory guidance is designed to support positioning before direct tool–tissue interaction and does not account for needle penetration or dynamic tissue deformation. Beyond localizing the needle, subretinal injection requires monitoring interaction-induced structural changes. In particular, successful fluid delivery produces a localized retinal detachment, commonly referred to as a subretinal bleb, whose formation and expansion must be monitored.

To address this gap, we introduce a real-time, deformation-aware sonification framework for subretinal injection. Unlike prior work relying on simulation of preoperative imaging dynamics~\cite{ruozzi2025biosonix} or focusing on pre-interaction auditory mappings~\cite{Matinfar2024OcularSA}, the proposed method encodes interaction-driven retinal deformation directly from intraoperative OCT using a physics-inspired acoustic model anchored to anatomical segmentations. This structured auditory representation enables unified encoding of needle depth, safety boundaries, and fluid-induced tissue deformation within a single extensible framework. In the context of subretinal injection, we focus on signals corresponding to ILM penetration marking retinal entry, proximity to the RPE as a critical safety boundary, and formation of the subretinal bleb confirming successful delivery. The system operates in real time and is compatible with interactive surgical workflows. The code is available at \url{https://github.com/luisdavid64/ioct-subretinal-sonification}.

We evaluate the proposed framework across simulated and ex vivo porcine iOCT sequences and assess its clinical relevance through expert feedback. In a controlled user study with 30 novices and 4 expert participants, the proposed sonification significantly improved identification of ILM penetration, RPE contact, and subretinal bleb formation compared to a parameter-mapping baseline. The largest gains were observed in deformation (bleb) detection, supporting the framework’s ability to convey dynamic interaction cues through audio alone and suggesting its potential to preserve visual focus during critical insertion phases.




\section{Related Work}

\textbf{Medical Sonification.}
Medical sonification has demonstrated great promise in augmenting clinical procedures through auditory displays \cite{Matinfar2022SonificationAA,Black2017ASO}. Early work applied to vitreoretinal surgery discretized the anatomical workspace into zones and used musical cues to convey tool-tip position and procedural state \cite{Matinfar_Soundtracks}. Subsequent approaches moved beyond musical or simple parameter mappings, introducing model-based sound synthesis frameworks that map imaging-derived signals to parameters of a physics-inspired acoustic model, thereby generating structured auditory feedback \cite{matinfar2025tissue}. However, these methods typically rely on preoperative imaging and therefore cannot accurately reflect dynamic intraoperative anatomical changes.

A more recent study has explored sonification of tissue deformation signals rather than tool localization alone~\cite{ruozzi2025biosonix}, aiming to provide richer insight into anatomical interaction. However, these approaches rely on simulated deformation derived from preoperative imaging, limiting their ability to reflect patient-specific and interaction-driven dynamics encountered intraoperatively.

OCT-specific sonification has been proposed for ERM peeling, mapping A-scans to a physics-inspired acoustic model to provide structural auditory feedback for identifying gaps between the ERM and ILM during membrane localization and peeling initiation~\cite{Matinfar2024OcularSA}. The auditory feedback primarily supports structural assessment prior to direct tool–tissue interaction and does not encode interaction-driven deformation dynamics relevant to subretinal injection.

\textbf{iOCT Processing for Subretinal Injection.}
In parallel, robotic and com\-puter-assisted subretinal injection has emerged as an active research area, driven by physiological limitations such as hand tremor and the difficulty of processing iOCT data in real time \cite{Arikan2024RealTimeDC}. This work has enabled key components for iOCT-guided insertion, including tool localization, real-time segmentation of retinal layers and instrument position, and extraction of tool-aligned B-scans from volumetric OCT data \cite{Arikan2024RealTimeDC,Dehghani2023RoboticNA,Zhou2025NeedleDA,Arikan2024TowardsMC}. Although these methods improve geometric estimation and visualization, effective intraoperative communication of this information remains challenging.

\section{Method}

\begin{figure}[htbp]
    \centering
    \includegraphics[width=0.95\linewidth]{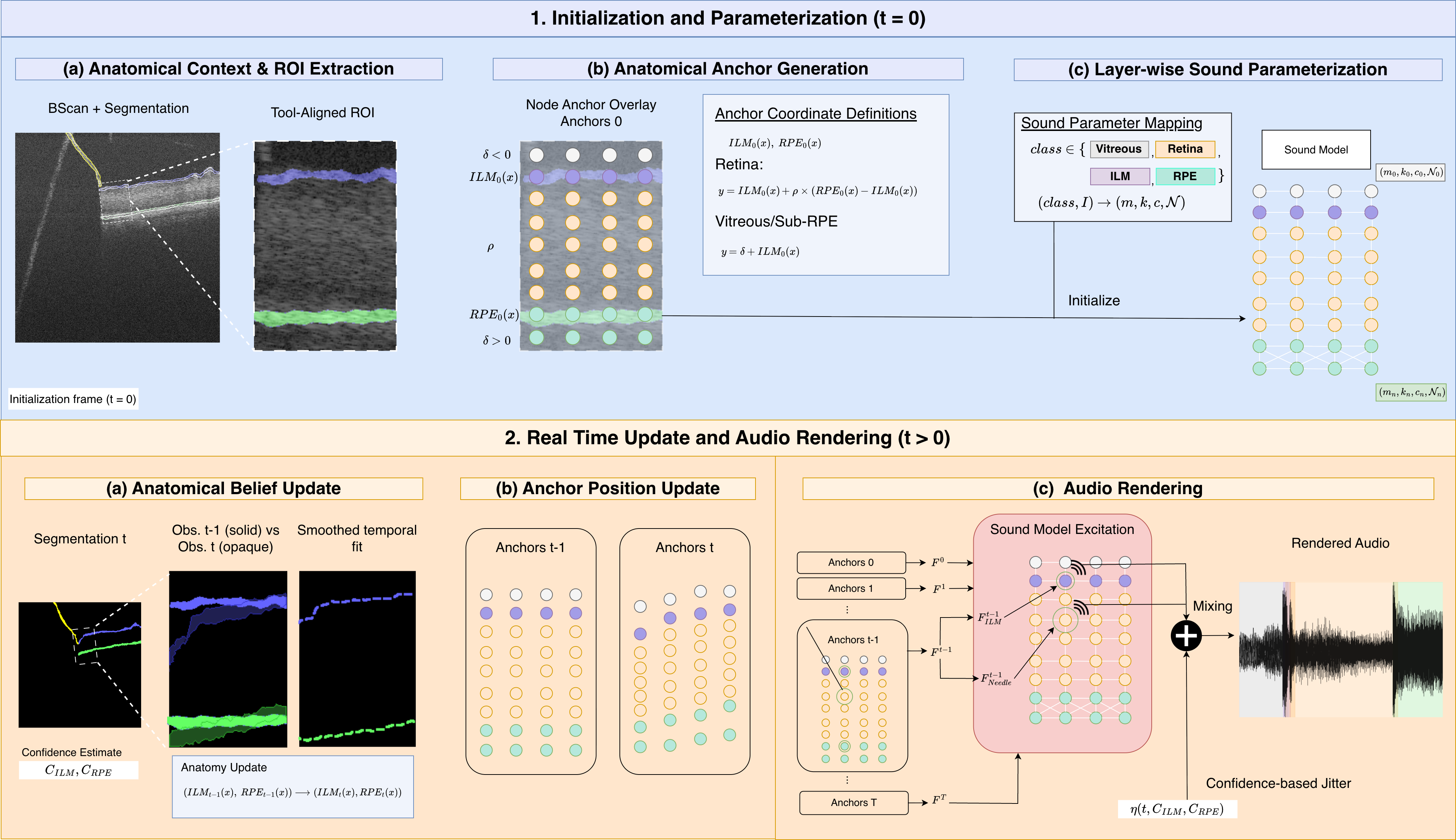}
    \caption{IOCT Sonification framework}
    \label{fig:method}
\end{figure}

Let $I_t \in \mathbb{R}^{H \times W}$ denote the incoming iOCT B-scan at frame $t$. 
Our goal is to map the image stream to a continuous audio signal $y(\tau)$ that reflects subretinal injection dynamics. 
The framework consists of (i) an initialization phase at $t = 0$, in which a physics-inspired sound model is instantiated and aligned with the segmented retinal anatomy, and (ii) a real-time update and audio rendering loop for $t > 0$.


\subsection{Initialization and Parameterization}

The underlying sound model is formulated as a two-dimensional mass--spring--damper system that produces audible resonances under external excitation \cite{Villeneuve2019MassInteractionPM}, following prior work in medical sonification \cite{ruozzi2025biosonix,Matinfar2024OcularSA}. 
The system consists of interconnected masses whose vibrational dynamics are governed by stiffness and damping parameters. 
During initialization, both the spatial configuration and physical properties are derived from geometric and semantic information extracted from the first segmented B-scan, thereby anchoring the model to the retinal anatomy. 
These parameters shape the spectral and resonant characteristics of the synthesized sound when nodes are excited, enabling contrastive feedback across tissue layers, implicitly encoding tool–tissue proximity, and intuitively reflecting deformations such as bleb formation.

\subsubsection{Anatomical Context and Tool-aligned ROI Extraction}

We process an initialization frame $I_0$ acquired during setup, where the needle shaft is visible in iOCT. A coarse segmentation $S_0$ of the ILM, RPE, and needle is obtained using a pre-trained U-Net, with low-confidence predictions suppressed as in~\cite{Dehghani2023RoboticNA}. From $S_0$, we estimate the dominant retinal orientation and compute a rotation matrix $R$ to align the tissue horizontally. The needle shaft is modeled by fitting a line $\ell$ to the segmented needle pixels using a robust Huber regressor \cite{Huber_1977_RME,Dehghani2023RoboticNA}. After applying $R$, a rotated rectangular region of interest (ROI) is extracted around the intersection of $\ell$ with the anatomical layers and extended to include the vitreous region above the ILM. If direct overlap is not detected (e.g., due to shadowing), the ROI is estimated from a local neighborhood search or from the needle trajectory alone.

\subsubsection{Anatomical Node Generation}

As shown in Fig.~\ref{fig:method}b, a uniform grid is overlaid on the ROI to define sound model nodes. Each node is assigned an anatomical class (vitreous, ILM, retina, RPE) via majority voting within its support, with ILM and RPE up-weighted to account for their thin structure.

Node positions are parameterized relative to the segmented $ILM_0$ and $RPE_0$ to ensure deformation-consistent alignment. For retinal nodes, axial position is defined by a normalized depth parameter $\rho \in [0,1]$:
\begin{equation}
y = \mathrm{ILM_0}(x) + \rho \big(\mathrm{RPE_0}(x) - \mathrm{ILM_0}(x)\big).
\end{equation}
Nodes outside the retina are defined by offsets to the nearest boundary, enabling runtime updates that preserve anatomical consistency of the anchors under deformation.

\subsubsection{Layer-wise Sound Parameterization}
\label{subsubsec:sound_param}
Our $m \times n$ anchor grid defines the masses of a physics-based sound model. 
To complete the mass--spring--damper system, we assign node-specific physical parameters and define the connectivity between neighboring masses to instantiate spring--damper elements.

Based on anatomical class labels and local intensity statistics within the ROI, we define a hand-crafted mapping
\begin{equation}
\mathbf{M}: (\text{class}, I) \rightarrow (m, k, d, \mathcal{N}),
\end{equation}
which assigns mass $m$, stiffness $k$, and damping $d$ to each node.

Nodes are connected in a regular lattice defined by a neighborhood order 
$\mathcal{N} \in \{1,2\}$, indicating first- or second-order ordinal proximity. 
First-order connectivity links each node to its immediate neighbors along the 
Cartesian axes, i.e., $(i\pm1,j)$ and $(i,j\pm1)$. 
Second-order connectivity extends coupling to the diagonal neighbors 
$(i\pm1,j\pm1)$.

For each neighboring pair $(i,j) \in \mathcal{N}$, spring stiffness and damping 
are defined symmetrically as
\begin{equation}
k_{ij} = \tfrac{1}{2}(k_i + k_j), 
\qquad 
d_{ij} = \tfrac{1}{2}(d_i + d_j).
\end{equation}

The resulting system determines the resonant and spectral characteristics of the synthesized sound.

\subsection{Real-time Anatomical Update (t>0)} \label{sec:update}

Following initialization, the sound model is updated online using the incoming iOCT B-scan stream. At fixed temporal intervals, each frame is segmented to obtain updated $ILM_t$, $RPE_t$, and needle estimates with associated confidence measures. To ensure robustness under partial occlusion and shadowing, ILM and RPE trajectories are modeled using cubic spline fitting with confidence-weighted regularization \cite{Boor1978APG,Unser1999SplinesAP} applied along the lateral axis. This lightweight temporal modeling enables smooth extrapolation across missing or low-confidence regions while remaining compatible with real-time processing constraints.

The updated anatomical estimates are then used to recompute node positions according to the relative parameterization introduced in the previous section. In particular, each node preserves its relative depth parameters $(\rho, \delta)$, while its absolute position is updated based on the current $ILM_t(x)$ and $RPE_t(x)$ geometry. This allows the anchors to follow tissue deformation over time.

\subsection{Excitation Protocol and Audio Rendering}

The sound model is driven by a set of excitation mechanisms designed to reflect both tool--tissue position and ongoing anatomical changes, while also conveying the reliability of the underlying visual estimates.

\subsubsection{Tool-driven Excitation.}
Excitations are derived from the estimated needle tip position. As the needle advances or interacts with tissue, forces are applied to the nearest node and propagated through the spring–damper lattice. Excitation magnitude is scaled by local stiffness~\cite{ruozzi2025biosonix}, ensuring anatomically consistent response across tissue types. This produces resonant events reflecting needle proximity and contact.

\subsubsection{Anatomy-driven Excitation.}

Surgical events such as tissue penetration or fluid injection induce localized deformation of the retinal layers, which manifests as differential axial motion between the ILM and RPE. Rather than directly measuring deformations in noisy iOCT scans, we estimate a coarse deformation by comparing temporal changes in the positions of these anatomical layers in the vicinity of the needle tip.

At each time step $t$, we define a lateral window $\mathcal{W}(x_t)$ centered at the current injection site (approximated by the needle tip position $x_t$). This focuses the analysis on tool–tissue and fluid injection–related interaction regions rather than global anatomical drift. Within this window, we compute the local ILM–RPE separation from the segmentations.
\begin{equation}
d_t(x) = RPE_t(x) - ILM_t(x), 
\quad x \in \mathcal{W}(x_t).
\end{equation}

The robust temporal change in separation is then defined as
\begin{equation}
\Delta d_t = 
\mathrm{P}_{95}\!\left(d_t(x) - d_{t-1}(x)\right),
\quad x \in \mathcal{W}(x_t),
\end{equation}
where $\mathrm{P}_{95}(\cdot)$ denotes the 95th percentile, 
used to suppress outliers due to segmentation noise. Finally, we define the deformation-driven excitation proxy
\begin{equation}
f_{ILM} = \min\!\big(2,\; \max(0,\; \Delta d_t)\big),
\end{equation}
which retains responsiveness to  positive separation increases corresponding to 
injection-induced retinal elevation while suppressing compression. 
The resulting signal is used to modulate excitation strength in the 
acoustic model. To prevent excessive excitation, $f_{ILM}$ is bounded 
to a predefined range.

\subsubsection{Confidence-modulated rendering.}

Confidence estimates are used to modulate a stochastic post-processing component in the sound synthesis $\eta(t,C_{ILM},C_{RPE})$, such that lower confidence results in increased temporal jitter. 

\section{Experiments and Result}

The sound model was implemented using the miPhysics library\footnote{\url{https://github.com/mi-creative/miPhysics\_Processing}} and motivate the physical parameter mapping on psychoacoustics $\mathbf{M}$ \cite{Serafin2011TheSH}.

\subsubsection{Qualitative Results}

\begin{figure}[t]
    \centering
    \includegraphics[width=0.9\linewidth]{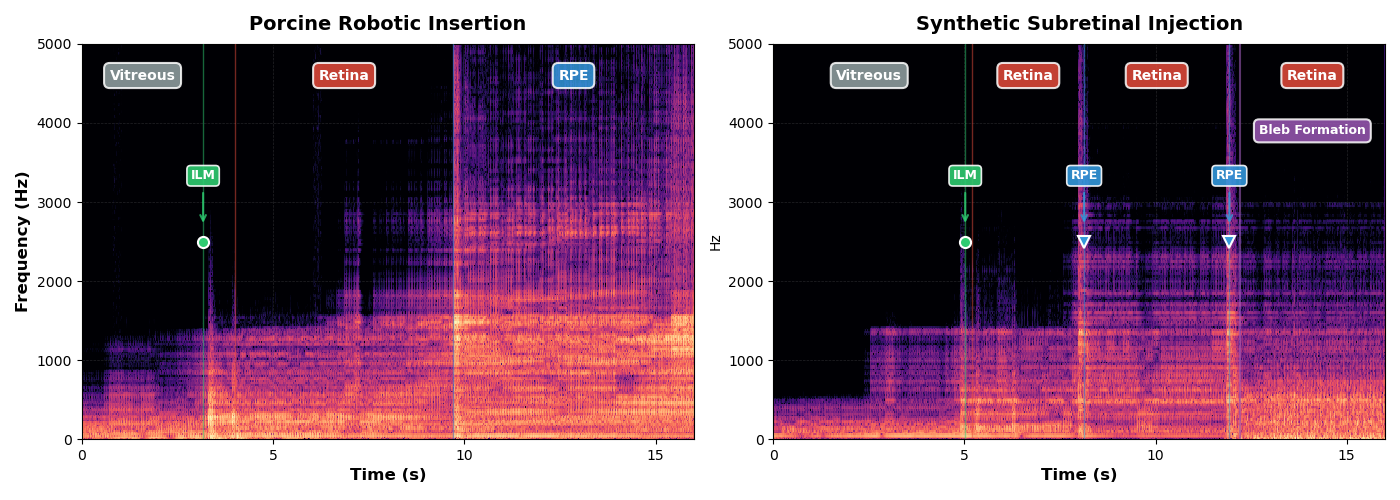}
    \caption{Spectrograms of sonified iOCT signals from (left) ex vivo porcine robotic insertion and (right) a synthetic subretinal injection sequence. In the synthetic example, bleb formation is characterized by a broadband increase in spectral energy.}
    \label{fig:porcien}
\end{figure}

We qualitatively evaluate the method on a publicly available porcine insertion~\cite{Pannek2024ExploringTN} and a representative synthetic sequence exhibiting bleb formation (Fig.~\ref{fig:porcien}). We provide corresponding supplementary videos (S01 and S02), and an example with vivid confidence modulation (S03). In both cases, detected events align closely with annotated ground truth, indicating accurate temporal localization.

\subsubsection{Real-time System -- Runtime Performance}  We demonstrate the full system operates in real time within an interactive simulated subretinal injection environment (supplementary video S04). We evaluate runtime on a CPU-bound system across 10 insertion sequences at $512\times512$ resolution. The full pipeline requires $27.8 \pm 1.5$ ms per frame (mean $\pm$ SD), corresponding to approximately 36 FPS. Segmentation and spline fitting dominated computation time, while all other components required $<1$ ms per frame.

\subsubsection{User Study}

We conducted a controlled within-subject study comparing the proposed sonification with a parameter-mapping baseline on realistic simulated subretinal injection sequences. Thirty-five participants were recruited; one was excluded (<25\% accuracy in both methods), resulting in 34 analyzed participants, from which 30 are novices and 4 are clinical experts. Each completed 20 trials (10 per method), identifying ILM contact, RPE contact, and bleb onset while rating confidence (1–5). Trials included single- and multi-event scenarios requiring correct temporal ordering. Method order was randomized, and trials were stratified by difficulty and randomized within strata. Differences ($\Delta$) are reported as proposed minus baseline and tested using the Wilcoxon signed-rank test. \textbf{The baseline} builds on pitch-based anatomical encoding strategies from prior surgical sonification frameworks~\cite{Matinfar_Soundtracks,Sonifeye_roodaki}, where static pitch represents anatomical zones and pulse-rate modulation proportional to ILM–RPE distance provides proximity feedback~\cite{Matinfar2022SonificationAA}. Parameterization follows established psychoacoustic principles~\cite{Fastl1990PsychoacousticsFA}, and prior work has demonstrated its effectiveness in supporting perceptual discrimination of anatomical regions. Bleb formation is conveyed indirectly through distance-dependent modulation rather than explicit deformation encoding.

\subsubsection{User Study Results}

\begin{figure}[t]
    \centering
    
    \begin{subfigure}[t]{0.45\linewidth}
        \centering
        \includegraphics[width=\linewidth]{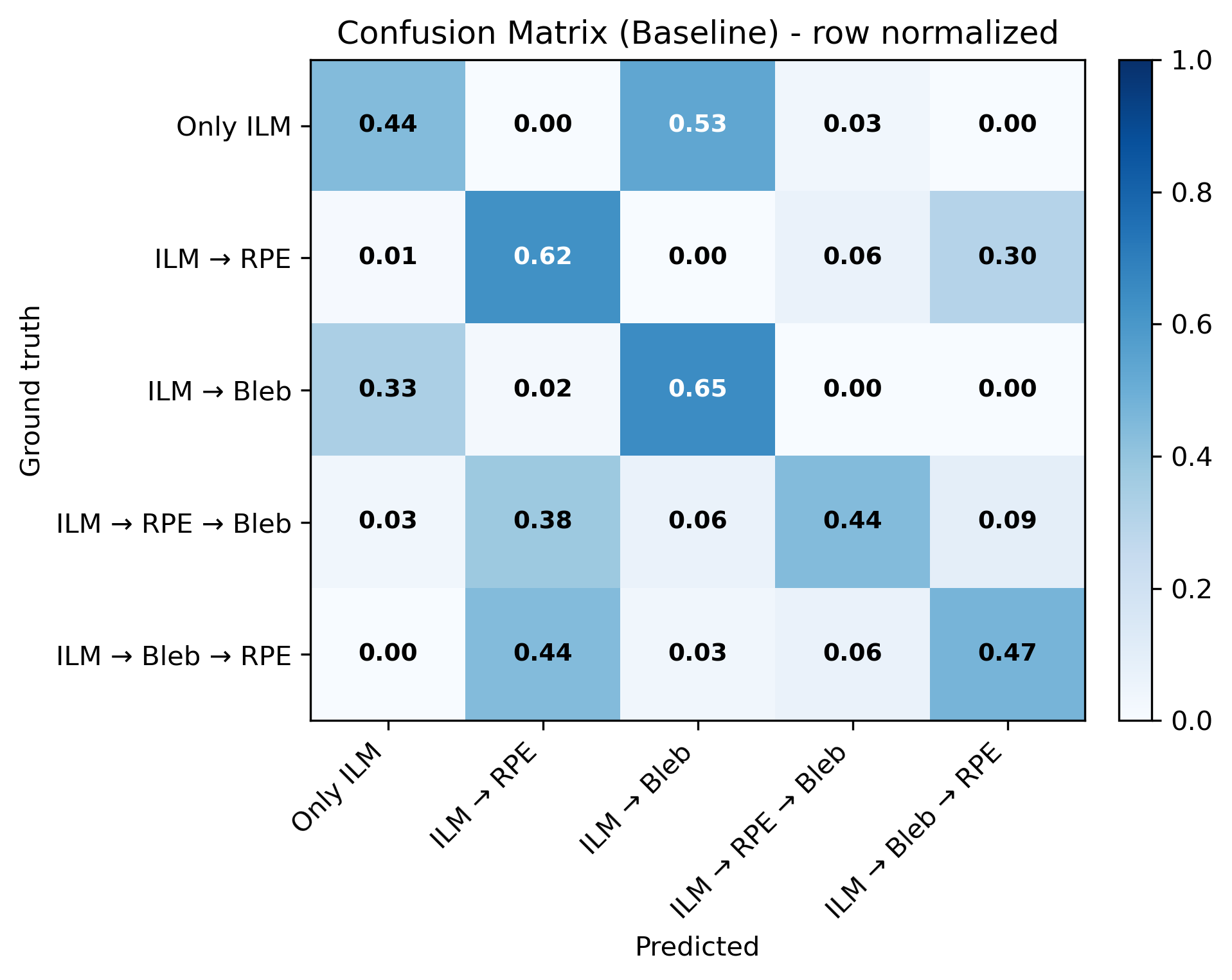}
        \caption{Baseline method}
        \label{fig:conf_baseline}
    \end{subfigure}
    \hfill
    \begin{subfigure}[t]{0.45\linewidth}
        \centering
        \includegraphics[width=\linewidth]{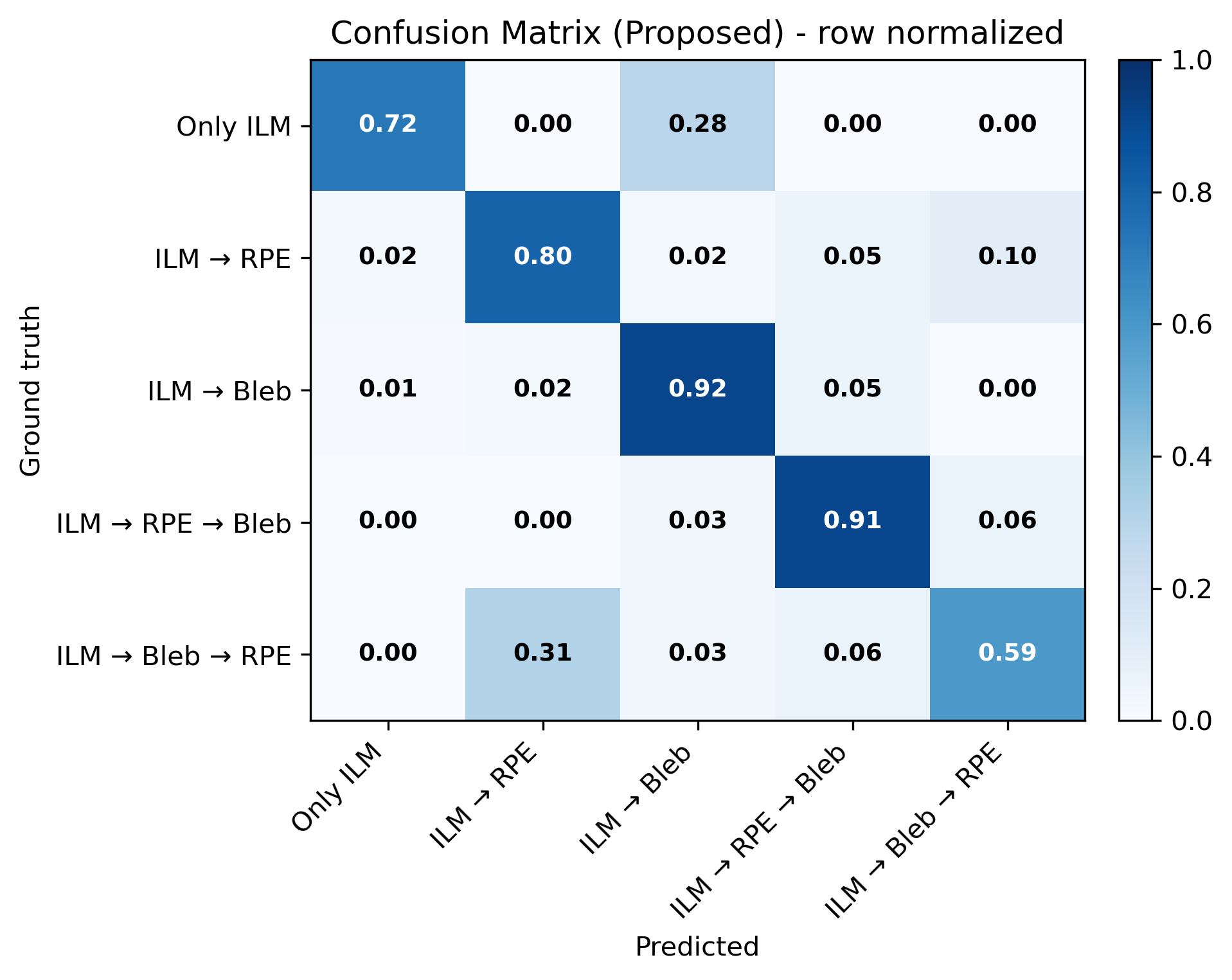}
        \caption{Proposed method}
        \label{fig:conf_proposed}
    \end{subfigure}
    
    \caption{Row-normalized confusion matrices (n=34). The baseline shows entanglement between contact-only and deformation states, while the proposed method improves separation of bleb-related events, reflected by stronger diagonal concentration.}
    \label{fig:confusion}
\end{figure}

The proposed sonification significantly improved overall event identification accuracy compared to the baseline (83.4\% vs.\ 60.6\%; $\Delta = 22.8$ percentage points, 95\% CI [15.7, 30.1], $p < 0.001$). Component-wise analysis showed that gains were primarily driven by improved bleb identification (63.6\% vs.\ 85.5\%; $\Delta = 21.8$ percentage points, $p < 0.001$), while ILM exhibited a minor numerical increase (+0.9\%) and RPE a small decrease (-1.6\%), neither statistically significant.  As illustrated in Fig.~\ref{fig:confusion}, the proposed method reduces structured confusions between contact-only and deformation states related to bleb formation. Participants reported significantly higher confidence on average under the proposed sonification (4.20 vs.\ 3.73; $\Delta = +0.47$, 95\% CI $[0.23, 0.71]$, $p < 0.001$), with the median confidence increasing from 4 to 5. Confidence was positively correlated with correctness in both conditions (Spearman $\rho = 0.27$ for both methods; both $p < 0.001$), indicating that subjective certainty reflected actual performance.

\subsubsection{Expert Feedback} Two vitreoretinal surgeons and two residents (n = 4) from the main cohort provided additional interview feedback.  In this group, recognition accuracy increased descriptively from 65\% (baseline) to 75\% (proposed), with comparable gains in bleb detection. Experts highlighted the method’s clinical relevance and its potential to reduce reliance on visual iOCT interpretation in favor of the optical microscope, potentially alleviating visual load \cite{SchuetziOCTCognitive}.

\section{Discussion and Conclusion}
We presented a real-time sonification framework for subretinal injection that converts iOCT B-scans into structured auditory feedback. Anchored in segmented retinal anatomy, a physics-inspired sound model encodes interaction-driven cues directly from intraoperative imaging rather than static proximity mappings. The user study demonstrated significantly improved identification of clinically relevant events, with the largest gains in bleb detection. These gains likely stem from the model’s physical formulation, which integrates deformation signals observable in iOCT, whereas the baseline relies on rigid parameterizations that do not accommodate such interaction cues. The framework supports real-time workflows and is extensible to additional interaction structures. Reported sound roughness by a few participants motivates future work in auditory and perceptual refinement \cite{Hermann2019DatadrivenAC} to further enhance usability.


    



%
%
%
\bibliographystyle{splncs04}
\bibliography{bibliography}
%

\end{document}